\documentclass[preprint,showpacs,preprintnumbers,amsmath,amssymb]{revtex4}
\usepackage{graphics,epsfig,subfigure}
\usepackage{color}

\begin{document}
\renewcommand{\baselinestretch}{1.3}
\newcommand\beq{\begin{equation}}
\newcommand\eeq{\end{equation}}
\newcommand\beqn{\begin{eqnarray}}
\newcommand\eeqn{\end{eqnarray}}
\newcommand\nn{\nonumber}
\newcommand\fc{\frac}
\newcommand\lt{\left}
\newcommand\rt{\right}
\newcommand\pt{\partial}

\title{Linear perturbations in Eddington-inspired Born-Infeld gravity}
\author{Ke Yang\footnote{yangke09@lzu.edu.cn},
        Xiao-Long Du\footnote{duxl11@lzu.edu.cn},
        and Yu-Xiao Liu\footnote{liuyx@lzu.edu.cn, corresponding author}}

\affiliation{Institute of Theoretical Physics, Lanzhou University, Lanzhou 730000, China.}

\begin{abstract}
We study the full linear perturbations of a homogeneous and isotropic spacetime in the Eddington-inspired Born-Infeld gravity. The stability of the perturbations are analyzed in the Eddington regime. We find that, for positive $\kappa$, the scalar modes are stable in the infinite wavelength limit ($k=0$) but unstable for $k\neq0$. The vector modes are stable and the tensor mode is unstable in the Eddington regime,  independent of the wave vector $k$. However, these modes are unstable and hence cause the instabilities for negative $\kappa$.
\end{abstract}

% \Keywords{ }



%\pacs{04.50.-h, 11.27.+d }

%11.10.Kk Field theories in dimensions other than four (see also 04.50.-h Higher-dimensional gravity and other theories of gravity; 04.60.Kz Lower dimensional models; minisuperspace models in general relativity and gravitation)

%04.50.Kd 	Modified theories of gravity

%04.50.-h Higher-dimensional gravity and other theories of gravity
%         (see also 11.25.Mj Compactification and four-dimensional models, 11.25.Uv D branes)

% 04.50.+h Gravity in more than four dimensions, Kaluza-Klein theory,
           % unified field theories, alternative theories of gravity
           %(see also 11.25.M Compactification and four-dimensional models), dilaton gravity
% 11.27.+d Extended classical solutions; cosmic strings,
           %domain walls, texture (see 98.80.C in cosmology)
%98.80.-k Cosmology
%\keywords{}

\maketitle


%%%%%%%%%%%%%%%%%%%%%%%%%%%%%%%%%%%%%%%%%%%%%%%%%%%%%%%%%%%%%%%%%%%%%%%%%%%%%%%%%%%%%%%

\section{Introduction}

A purely affine theory of gravity was proposed by Eddington in 1924 \cite{Eddington1924,Schrodinger1950}. The theory is totally equivalent to general relativity with a cosmological constant. Since matter is not included, it is not a complete theory. Recently, a new alternative theory called Eddington-inspired Born-Infeld (EiBI) gravity put forward by Banados and Ferreira \cite{Banados2010} has drawn much attention (see also Refs. \cite{Deser1998,Vollick2004}). They extended Eddington's theory with matter included in a conventional way. This theory is totally equivalent to generally relativity in the vacuum but deviates from it when matter is present. As the most attractive feature, the homogeneous and isotropic cosmology seem to be singularity free in this theory \cite{Banados2010,Scargill2012}. Further research also shows that singularities may be prevented during gravitational collapse \cite{Pani2011}. Reference \cite{Avelino2012a} argued that the EiBI theory may be a viable alternative to the inflationary paradigm to solve the fundamental problems of the standard cosmological model. On the other hand, some pathologies of the EiBI theory were also pointed out in recent works. The studies in Refs. \cite{Escamilla-Rivera2012,Avelino2012a} showed that the tensor perturbation is linearly unstable deep in the Eddington regime. And Ref. \cite{Pani2012a} showed that there exist curvature singularities at the surface of polytropic stars and the unacceptable Newtonian limit in this theory.
The authors in Ref. \cite{Bouhmadi-Lopez2013} found that the big rip singularity is unavoidable in the EiBI phantom model. More relevant studies can be seen in Refs. \cite{Casanellas2012,Avelino2012,Pani2012,Delsate2012,Liu2012a,Avelino2012b,Sham2012,Cho2012,Jana2013,Cho2013,
Bouhmadi-Lopez2013,Sham2013,Harko2013,Cho2013a,Harko2013a}.

Research shows that a  homogeneous and isotropic Universe will reach a maximum density at an early time. However, the stability of the evolution near this maximum density is an important problem that should be considered. The instability of linear tensor perturbations was investigated in Refs. \cite{Escamilla-Rivera2012,Avelino2012a}. In this paper, we investigate the stability problem of full linear perturbations, including the scalar, vector, and tensor perturbations of a homogeneous and isotropic spacetime in the EiBI theory. We find that the tensor mode diverges and causes an instability when approaching the maximum density for both $\kappa>0$ and $\kappa<0$. This work overlaps the result of tensor perturbation in Refs. \cite{Escamilla-Rivera2012,Avelino2012a}. However, the vector modes do not diverge and hence preserve stabilities for $\kappa>0$, while they diverge and cause instabilities for $\kappa<0$. For the scalar modes, they are stable only in the case of $k=0$ and $\kappa>0$.

The paper is organized as follows. In Sec. \ref{The_linearized_field_equations}, we derive the linearized equations of motion of the theory. In Sec. \ref{The_stability_of_the_perturbations}, the stability of scalar, vector, and tensor modes is studied. Finally, conclusions are presented.

\section{Linearized field equations} \label{The_linearized_field_equations}

Working in Planck units $c=8\pi G=1$, the action for the EiBI theory is given by \cite{Banados2010}
\beq
S=\frac{2}{\kappa}\int{d^4 x\lt[\sqrt{|g_{\mu\nu}+\kappa R_{\mu\nu}(\Gamma)|}-\lambda\sqrt{g} \rt]}+S_{M},
\label{EiBI_Action}
\eeq
where $R_{\mu\nu}(\Gamma)$ represents the symmetric part of the Ricci tensor and $\lambda=1+\kappa\Lambda$ is a nonvanishing constant. When $g\gg\kappa R$, the action reproduces the Einstein-Hilbert action with cosmological constant $\Lambda$. On the other hand, when $\kappa R\gg g$, the action approximates to Eddington's. Since the metric and the connection are treated as independent variables, the equations of motion are given by \cite{Banados2010}

\beqn
\sqrt{q}q^{\mu\nu}&=&\lambda \sqrt{g}g^{\mu\nu}-\kappa\sqrt{g}T^{\mu\nu},\label{Field_Eq_1}\\
q_{\mu\nu}&=&g_{\mu\nu}+\kappa R_{\mu\nu},\label{Field_Eq_2}
\eeqn
where $q_{\mu\nu}$ is the auxiliary metric compatible to the connection, i.e., $\Gamma^{\lambda}_{\mu\nu}=\fc{1}{2}q^{\lambda\sigma}(q_{\sigma\mu,\nu}+q_{\sigma\nu,\mu}-q_{\mu\nu,\sigma})$, and $q^{\mu\nu}$ is the inverse of $q_{\mu\nu}$.

Now, we consider a perturbed homogeneous and isotropic spacetime with these two metrics,
\beqn
ds^2&=&g_{\mu\nu}dx^{\mu}dx^{\nu}\nonumber \\
    &=&(-1+h_{00})dt^2+a^2(\delta_{ij}+h_{ij})dx^{i}dx^{j}+2h_{0i}dtdx^{i},\\
d\tilde{s}^2&=&q_{\mu\nu}dx^{\mu}dx^{\nu}\nonumber \\
    &=&X^2(-1+\gamma_{00})dt^2+a^2Y^2(\delta_{ij}+\gamma_{ij})dx^{i}dx^{j}+2Y^2\gamma_{0i}dtdx^{i},
\label{Per_Metric}
\eeqn
where $a(t), X(t)$, and $Y(t)$ are background quantities depending solely on the time and $h(x), \gamma(x)$ are the perturbations involving both the space and time.

The energy-momentum conservation equation $\nabla_{\mu}T^{\mu}_{\nu}=0$ for matter fields in the EiBI theory is held as in general relativity, where the covariant derivative refers to the metric $g_{\mu\nu}$. For the perfect fluid, the energy-momentum tensor is given by $T^{\mu\nu}=Pg^{\mu\nu}+(P+\rho)u^{\mu}u^{\nu}$
with $u^{\mu}$ the four-velocity of the static observer $u^{\mu}=(1,0,0,0)$. After perturbing the fluid, i.e., $\tilde P=P+\delta P$, $\tilde \rho=P+\delta \rho$ and $\tilde u^{\mu}=u^{\mu}+\delta u^{\mu}$, the first-order perturbed energy-momentum tensor is given by
\beqn
\delta T^{00}&=&\delta \rho+\rho h_{00},\\
\delta T^{i0}&=&-a^{-2}\rho h_{0i}+a^{-2}(P+\rho)\delta u_{i},\\
\delta T^{ij}&=&a^{-2}\delta P \delta_{ij}-a^{-2}P h_{ij}.
\eeqn
Substituting these metrics and energy-momentum tensor into the field equations (\ref{Field_Eq_1}) and (\ref{Field_Eq_2}), the zeroth-order equation of Eq. (\ref{Field_Eq_1}) gives
\begin{subequations}\label{BG_EOM_1}
\beqn
{Y^3}/{X}&=&\lambda+\kappa\rho,\\
XY&=&\lambda-\kappa{}P,
\eeqn
\end{subequations}
and the zeroth-order equation of Eq. (\ref{Field_Eq_2}) gives
\begin{subequations}\label{BG_EOM_2}
\beqn
 X^2 &=& 1 + 3\kappa \left[\frac{{\ddot a}}{a} + \frac{{\ddot Y}}{Y} - \frac{{\dot a}}{a}\frac{{\dot X}}{X} + 2\frac{{\dot a}}{a}\frac{{\dot Y}}{Y} - \frac{{\dot X}}{X}\frac{{\dot Y}}{Y}\right],\\
 Y^2 &=& 1 + \kappa \frac{{{Y^2}}}{{{X^2}}}\left(\frac{{\ddot a}}{a}
         + 2\frac{{{{\dot a}^2}}}{{{a^2}}}
         - \frac{{\dot a}}{a}\frac{{\dot X}}{X}
         + 6\frac{{\dot a}}{a}\frac{{\dot Y}}{Y}\right. \nonumber \\
      && \left. - \frac{{\dot X}}{X}\frac{{\dot Y}}{Y}
         + \frac{{\ddot Y}}{Y} + 2\frac{{{{\dot Y}^2}}}{{{Y^2}}}\right).
\eeqn
\end{subequations}
Equations (\ref{BG_EOM_2}) will be used to simplify the first-order perturbed equations.

After some algebra, the perturbed auxiliary metric is read from the first-order equation of Eq. (\ref{Field_Eq_1}),
\begin{subequations}\label{Perturbations_of_Aux_Metric}
\beqn
\gamma_{00}&=&h_{00}+\fc{\kappa \delta \rho}{2(\lambda  + \kappa \rho) }+\fc{{3\kappa \delta P}}{2({\lambda  - \kappa P})},\\
{\gamma _{0i}} &=& {h_{0i}} - \fc{ (P + \rho )}{(\lambda+\kappa\rho)}\kappa\delta {u_i},\\
\gamma_{ij}&=&h_{ij}+\fc{\kappa \delta \rho{\delta _{ij}}}{2(\lambda  + \kappa \rho) }-\fc{{\kappa \delta P}{\delta _{ij}}}{2({\lambda  - \kappa P})}.
\eeqn
\end{subequations}
Here, we note that the transverse-traceless parts of $h_{ij}$ and $\gamma_{ij}$ are interestingly identical to each other, as noticed in Refs. \cite{Escamilla-Rivera2012,Liu2012a}.

With the perturbed auxiliary metric (\ref{Perturbations_of_Aux_Metric}), the first-order equation of Eq. (\ref{Field_Eq_2}) will give  three dynamical equations for the perturbation modes $h_{00}$, $h_{i0}$, and $h_{ij}$. On the other hand, after perturbing the energy-momentum conservation equation $\nabla_{\mu}T^{\mu}_{\nu}=0$, we will obtain another two dynamical equations. But they are not independent and can be derived from the former three equations.

Further, to study the evolution of perturbations, it is convenient to decompose the perturbations into scalar, transverse vector and transverse-traceless tensor modes, which are not coupled to each other by the field equations or conservation equations \cite{Weinberg2008}. Here, the perturbations are decomposed as
\beqn
h_{00}&=&-E,~~~~~ h_{i0}=\pt_i{F}+G_i, \\
h_{ij}&=&A\delta_{ij}+\pt_i\pt_j B+\pt_j C_i +\pt_i C_j+D_{ij},\\
\delta u_{i}&=&\pt_{i}\delta u+\delta U_i,
\eeqn
where $\pt_i C_i=\pt_i G_i=\pt_i \delta U_i=0$, $\pt_{i}D_{ij}=0$, and $D_{ii}=0$. Thus, the perturbed equations involve seven scalar modes $E, F, A, B, \delta\rho, \delta P, \delta u$, three transverse vector modes $C_i, G_i, \delta U_i$, and one transverse-traceless tensor mode $D_{ij}$.

\subsection{Scalar modes}

The $00$ component of the perturbed equation (\ref{Field_Eq_2}) gives
\begin{widetext}
\beqn
\!&\!\!&\!\frac{1}{2}\frac{{{X^2}}}{{{Y^2}}}{a^{ - 2}}{\nabla ^2}E + 3(\frac{{\ddot a}}{a} + \frac{{\ddot Y}}{Y} - \frac{{\dot a}}{a}\frac{{\dot X}}{X} + 2\frac{{\dot a}}{a}\frac{{\dot Y}}{Y} - \frac{{\dot X}}{X}\frac{{\dot Y}}{Y})E + \frac{3}{2}(\frac{{\dot a}}{a} + \frac{{\dot Y}}{Y})\dot E - \frac{1}{2}(3\ddot A + {\nabla ^2}\ddot B) \nn \\
  \!&\!\!&\!  - (\frac{{\dot a}}{a} - \frac{1}{2}\frac{{\dot X}}{X} + \frac{{\dot Y}}{Y})(3\dot A + {\nabla ^2}\dot B) + {a^{ - 2}}{\nabla ^2}\dot F + {a^{ - 2}}(2\frac{{\dot Y}}{Y} - \frac{{\dot X}}{X}){\nabla ^2}F - \frac{\kappa }{4}{a^{ - 2}}\frac{{{X^2}}}{{{Y^2}}}\frac{{{\nabla ^2}\delta \rho }}{{\lambda  + \kappa \rho }} \nn \\
  \!&\!\!&\!  - \frac{{3\kappa }}{4}{a^{ - 2}}\frac{{{X^2}}}{{{Y^2}}}\frac{{{\nabla ^2}\delta P}}{{\lambda  - \kappa P}} - \frac{{3\kappa }}{4}{\partial _0}{\partial _0}\frac{{\delta \rho }}{{\lambda  + \kappa \rho }} - \frac{{3\kappa }}{4}(3\frac{{\dot a}}{a} + 3\frac{{\dot Y}}{Y} - \frac{{\dot X}}{X}){\partial _0}\frac{{\delta \rho }}{{\lambda  + \kappa \rho }}+ \frac{{3\kappa }}{4}{\partial _0}{\partial _0}\frac{{\delta P}}{{\lambda  - \kappa P}} \nn \\
   \!&\!\!&\! - \frac{{3\kappa }}{4}(\frac{{\dot a}}{a} + \frac{{\dot Y}}{Y} + \frac{{\dot X}}{X}){\partial _0}\frac{{\delta P}}{{\lambda  - \kappa P}}  - \frac{1}{2}[1 + 3\kappa (\frac{{\ddot a}}{a} + \frac{{\ddot Y}}{Y} - \frac{{\dot a}}{a}\frac{{\dot X}}{X} + 2\frac{{\dot a}}{a}\frac{{\dot Y}}{Y} - \frac{{\dot X}}{X}\frac{{\dot Y}}{Y})](\frac{{\delta \rho }}{{\lambda  + \kappa \rho }} + \frac{{3\delta P}}{{\lambda  - \kappa P}}) \nn \\
   \!&\!\!&\!- \kappa {a^{ - 2}}\pt_0(\frac{{P + \rho }}{{\lambda  + \kappa \rho }}{\nabla ^2}\delta  u ) - \kappa {a^{ - 2}}(2\frac{{\dot Y}}{Y} - \frac{{\dot X}}{X})\frac{{P + \rho }}{{\lambda  + \kappa \rho }}{\nabla ^2}\delta u = 0.
\label{Per_Scalar_Eq_1}
\eeqn
The part with the form $\pt_i S$ (where $S$ is any scalar) of the $i0$ component of the perturbed equation (\ref{Field_Eq_2}) gives
\beqn
\!&\!\!&\!(\frac{{\dot a}}{a} + \frac{{\dot Y}}{Y})E - \dot A - \frac{\kappa }{2}{\partial _0}\frac{{\delta \rho }}{{\lambda  + \kappa \rho }} + \frac{\kappa }{2}{\partial _0}\frac{{\delta P}}{{\lambda  - \kappa P}}
 %\nn\\
 %\!&\!\!&\!
 - \frac{\kappa }{2}(\frac{{\dot a}}{a} + \frac{{\dot Y}}{Y})(\frac{{\delta \rho }}{{\lambda  + \kappa \rho }} + \frac{{3\delta P}}{{\lambda  - \kappa P}})
 + \frac{{P + \rho }}{{\lambda  + \kappa \rho }}\delta u = 0.
\label{Per_Scalar_Eq_2}
\eeqn
The part of the $ij$ component of the perturbed equation (\ref{Field_Eq_2}) proportional to $\delta_{ij}$ gives
\beqn
\!&\!\!&\!- \frac{1}{2}\frac{{{a^2}{Y^2}}}{{{X^2}}}(\frac{{\dot a}}{a} + \frac{{\dot Y}}{Y})\dot E - \frac{{{a^2}{Y^2}}}{{{X^2}}}(\frac{{\ddot a}}{a} + \frac{{\ddot Y}}{Y} + 2\frac{{{{\dot a}^2}}}{{{a^2}}} - \frac{{\dot a}}{a}\frac{{\dot X}}{X} + 6\frac{{\dot a}}{a}\frac{{\dot Y}}{Y} + 2\frac{{{{\dot Y}^2}}}{{{Y^2}}} - \frac{{\dot X}}{X}\frac{{\dot Y}}{Y})E + \frac{1}{2}\frac{{{a^2}{Y^2}}}{{{X^2}}}\ddot A\nn \\
  \!&\!\!&\! - \frac{1}{2}{\nabla ^2}A- \frac{{{Y^2}}}{{{X^2}}}(\frac{{\dot a}}{a} + \frac{{\dot Y}}{Y}){\nabla ^2}F + \frac{1}{2}\frac{{{a^2}{Y^2}}}{{{X^2}}}(3\frac{{\dot a}}{a} + 3\frac{{\dot Y}}{Y} - \frac{{\dot X}}{X})\dot A + \frac{1}{2}\frac{{{a^2}{Y^2}}}{{{X^2}}}(\frac{{\dot a}}{a} + \frac{{\dot Y}}{Y})(3\dot A + {\nabla ^2}\dot B)\nn \\
   \!&\!\!&\! + \frac{\kappa }{4}\frac{{{a^2}{Y^2}}}{{{X^2}}}{\partial _0}{\partial _0}\frac{{\delta \rho }}{{\lambda  + \kappa \rho }}  - \frac{\kappa }{4}\frac{{{a^2}{Y^2}}}{{{X^2}}}{\partial _0}{\partial _0}\frac{{\delta P}}{{\lambda  - \kappa P}}  - \frac{\kappa }{4}(\frac{{{\nabla ^2}\delta \rho }}{{\lambda  + \kappa \rho }} - \frac{{{\nabla ^2}\delta P}}{{\lambda  - \kappa P}}) - \frac{1}{2}{a^2}(\frac{{\delta \rho }}{{\lambda  + \kappa \rho }} - \frac{{\delta P}}{{\lambda  - \kappa P}})\nn \\
   \!&\!\!&\!  + \frac{\kappa }{2}\frac{{{a^2}{Y^2}}}{{{X^2}}}(\frac{{\ddot a}}{a} + \frac{{\ddot Y}}{Y} + 2\frac{{{{\dot a}^2}}}{{{a^2}}} - \frac{{\dot a}}{a}\frac{{\dot X}}{X} + 6\frac{{\dot a}}{a}\frac{{\dot Y}}{Y} + 2\frac{{{{\dot Y}^2}}}{{{Y^2}}} - \frac{{\dot X}}{X}\frac{{\dot Y}}{Y})(\frac{{\delta \rho }}{{\lambda  + \kappa \rho }} + \frac{{3\delta P}}{{\lambda  - \kappa P}}) \nn \\
   \!&\!\!&\!  + \frac{\kappa }{4}\frac{{{a^2}{Y^2}}}{{{X^2}}}(7\frac{{\dot a}}{a} + 7\frac{{\dot Y}}{Y} - \frac{{\dot X}}{X}){\partial _0}\frac{{\delta \rho }}{{\lambda  + \kappa \rho }} - \frac{\kappa }{4}\frac{{{a^2}{Y^2}}}{{{X^2}}}(3\frac{{\dot a}}{a} + 3\frac{{\dot Y}}{Y} - \frac{{\dot X}}{X}){\partial _0}\frac{{\delta P}}{{\lambda  - \kappa P}}\nn \\
   \!&\!\!&\! + \kappa \frac{{{Y^2}}}{{{X^2}}}(\frac{{\dot a}}{a} + \frac{{\dot Y}}{Y})\frac{{P + \rho }}{{\lambda  + \kappa \rho }}{\nabla ^2}\delta u=0.
\label{Per_Scalar_Eq_3}
\eeqn
\end{widetext}
The part with the form $\pt_i\pt_j S$ of the $ij$ component of the perturbed equation (\ref{Field_Eq_2}) gives
\beqn
\!&\!\!&\!  - \frac{1}{2}E - \frac{1}{2}A + \frac{1}{2}\frac{{{a^2}{Y^2}}}{{{X^2}}}\ddot B + \frac{1}{2}\frac{{{a^2}{Y^2}}}{{{X^2}}}(3\frac{{\dot a}}{a} + 3\frac{{\dot Y}}{Y} - \frac{{\dot X}}{X})\dot B \nn \\
\!&\!\!&\! - \frac{{{Y^2}}}{{{X^2}}}\dot F - \frac{{{Y^2}}}{{{X^2}}}(\frac{{\dot a}}{a} - \frac{{\dot X}}{X} + 3\frac{{\dot Y}}{Y})F
   + \kappa \frac{{{Y^2}}}{{{X^2}}}{\partial _0}(\frac{{P + \rho }}{{\lambda  + \kappa \rho }}\delta u) \nn \\
\!&\!\!&\! + \kappa \frac{{\delta P}}{{\lambda  - \kappa P}}  + \kappa \frac{{{Y^2}}}{{{X^2}}}(\frac{{\dot a}}{a} - \frac{{\dot X}}{X} + 3\frac{{\dot Y}}{Y})\frac{{P + \rho }}{{\lambda  + \kappa \rho }}\delta u = 0.
\label{Per_Scalar_Eq_4}
\eeqn
The $0$ component of the perturbed conservation equation gives
\beqn
&&\delta \dot \rho  + 3\frac{{\dot a}}{a}(\delta \rho  + \delta P) + \frac{1}{2}(P + \rho )(3\dot A + {\nabla ^2}\dot B) \nn \\
\!&\!\!&\!- {a^{ - 2}}(P + \rho ){\nabla ^2}(F - \delta u)= 0.
\label{Per_Scalar_Eq_5}
\eeqn
The part with the form $\pt_i S$ of the $i$ component of the perturbed conservation equation gives
\beqn
\delta P + \frac{1}{2}(P + \rho )E + \pt_0[(P + \rho )\delta  u] + 3\frac{{\dot a}}{a}(P + \rho )\delta u=0.
\label{Per_Scalar_Eq_6}
\eeqn

\subsection{Vector modes}

The part with the form $V_i$ (where $V_i$ is any vector satisfying $\pt_i V_i=0$) of the $i0$ component of the perturbed equation (\ref{Field_Eq_2}) gives
\beqn
   \frac{1}{2}{\nabla ^2}{{\dot C}_i}
 - \frac{1}{2a^2}{\nabla ^2}{G_i} + \frac{\kappa }{2a^2}\frac{{P + \rho }}{{\lambda  + \kappa \rho }}{\nabla ^2}\delta {U_i} + \frac{{P + \rho }}{{\lambda  + \kappa \rho }}\delta {U_i} = 0.
\label{Per_vector_Eq_1}
\eeqn
The part with the form $\pt_j V_i $ of the $ij$ component of the perturbed equation (\ref{Field_Eq_2}) gives
\beqn
\!&\!\!&\! a^2{{\ddot C}_i} + a^2(3\frac{{\dot a}}{a} + 3\frac{{\dot Y}}{Y} - \frac{{\dot X}}{X}){{\dot C}_i} - {{\dot G}_i} - (\frac{{\dot a}}{a} - \frac{{\dot X}}{X} + 3\frac{{\dot Y}}{Y}){G_i}  \nn \\
\!&\!\!&\! + \kappa{\partial _0}(\frac{{P + \rho }}{{\lambda  + \kappa \rho }}\delta {U_i}) + \kappa(\frac{{\dot a}}{a} - \frac{{\dot X}}{X} + 3\frac{{\dot Y}}{Y})\frac{{P + \rho }}{{\lambda  + \kappa \rho }}\delta {U_i}=0.
\label{Per_vector_Eq_2}
\eeqn
The part of $i$ component of the perturbed conservation equation of the form $V_i$ gives
\beqn
\pt_0[(P + \rho )\delta  U_i] + 3\frac{{\dot a}}{a}(P + \rho )\delta {U_i} = 0.
\label{Per_vector_Eq_3}
\eeqn

\subsection{Tensor mode}

The part with the form of a transverse-traceless tensor of the $ij$ component of the perturbed equation (\ref{Field_Eq_2}) is
\beq
 {{\ddot D}_{ij}} + (3\frac{{\dot a}}{a} + 3\frac{{\dot Y}}{Y} - \frac{{\dot X}}{X}){{\dot D}_{ij}}- \frac{X^2}{{a^2}{Y^2}}{\nabla ^2}{D_{ij}} =0.
\label{Per_tensor_Eq}
\eeq

Here, we note that there are seven scalar modes, but only four of six equations are independent, and three transverse vector modes, but only two of three equations are independent. However, since the field equations (\ref{Field_Eq_1}) and (\ref{Field_Eq_2}) are invariant under a coordinate transformation $x^{\mu}\rightarrow x'^{\mu}$, these perturbed modes are not all physical \cite{Weinberg2008}. After eliminating the gauge degrees of freedom, one can remove two scalar modes and one vector mode. The transverse-traceless tensor mode is gauge invariant. So with the state equation $\tilde P=\omega\tilde\rho$, then $\delta P=\omega\delta\rho$, all the remaining perturbed modes are solvable.

\section{Stability of the perturbations} \label{The_stability_of_the_perturbations}

In this section we study the approximative evolution of the perturbed modes in the Eddington regime, in which the dominant constituent of the Universe is the highly relativistic ideal gas and the cosmological constant can be neglected. Thus, the state parameter $\omega=1/3$ and $\lambda=1$. For scalar modes, we work in the Newtonian gauge, i.e., we set $B=F=0$ in the linear perturbed equations, and for vector modes, we fix the gauge freedom to eliminate the mode $C_i$.

\subsection{$\kappa>0$}

For the case $\kappa>0$, the approximate background solution near the maximum density ($t\rightarrow-\infty$) is given by \cite{Escamilla-Rivera2012,Scargill2012}
\begin{subequations}\label{Background_Solution_1}
\beqn
a&=&a_B(1+e^{b(t-t_0)}),\\
X&=&2e^{\frac{3}{4}b(t-t_0)},\\
Y&=&2e^{\frac{1}{4}b(t-t_0)},
\eeqn
\end{subequations}
where $b=({8}/{3\kappa})^{\fc{1}{2}}$.
To get the evolution of the scalar modes, we use the four simpler equations, Eqs. (\ref{Per_Scalar_Eq_2}), (\ref{Per_Scalar_Eq_4}), (\ref{Per_Scalar_Eq_5}), and (\ref{Per_Scalar_Eq_6}). For transverse vector modes, we use Eqs. (\ref{Per_vector_Eq_2}) and (\ref{Per_vector_Eq_3}). Equation (\ref{Per_tensor_Eq}) is used to solve the transverse-traceless tensor mode. In calculation, we replace the Laplace operator with $k^2$, where $k$ is wave vector. With this solution (\ref{Background_Solution_1}) and zeroth-order field equation (\ref{BG_EOM_1}), the corresponding equations can be approximately simplified and solved in the limit $t\rightarrow-\infty$. Finally, we arrive at the solutions
\begin{subequations}\label{Solution_scalar}
\beqn
A&\simeq&{c_1}+{c_2} k^2 t +{c_3}{e^{\fc{7}{4}b(t - {t_0})}},\\
E&\simeq&({c_4}+{c_5}k^2t){e^{b(t - {t_0})}},\\
\delta\rho&\simeq&({c_6}+{c_7}k^2t){e^{b(t - {t_0})}},\\
\delta u &\simeq& {c_8}+c_9{e^{\fc{7}{4}b(t - {t_0})}}.
\eeqn
\end{subequations}
\beqn
{G_i} &\simeq& c_{10}, ~~~~
\delta {U_i} \simeq  c_{11}. \label{Solution_vector} \\
{D_{ij}} &\simeq& c_{12}t + c_{13},
\label{Solution_tensor}
\eeqn
where $c_i$ are functions of wave vector. In the infinite wavelength limit ($k=0$), the scalar modes do not diverge when the Universe approaches the maximum density. But for $k\neq0$, the scalar modes are linearly divergent. So we conclude that the scalar perturbations are stable for $k=0$ modes but unstable for $k\neq0$ modes in the Eddington regime. The tensor mode blows up as $t\rightarrow-\infty$, and it causes an instability as claimed in Ref. \cite{Escamilla-Rivera2012}. However, the vector perturbations are stable in the Eddington regime.

\subsection{$\kappa<0$}

For $\kappa<0$, the approximate background solution near the maximum density ($t\rightarrow0$) is given by \cite{Escamilla-Rivera2012,Scargill2012}
\begin{subequations}\label{Background_Solution_2}
\beqn
a&=&a_B[1-\frac{2}{3\kappa}|t|^2],\\
X&=&U^{\fc{1}{2}}=\frac{2}{\sqrt{3}}\sqrt[4]{-\frac{\kappa}{2}}|t|^{-\fc{1}{2}},\\
Y&=&V^{\fc{1}{2}}=\frac{2}{\sqrt{3}}\sqrt[4]{-\frac{2}{\kappa}}|t|^{\fc{1}{2}}.
\eeqn
\end{subequations}
With similar calculation, the perturbed modes near the maximum density are approximately given by
\begin{subequations}\label{Solution_scalar_2}
\beqn
A&\simeq&C_1|t|^{\frac{3}{2}}+C_2|t|^{\varepsilon},\\
E&\simeq&C_3|t|^{-\frac{1}{2}}+C_4|t|^{-2+\varepsilon},\\
\delta \rho  &\simeq&{C_5}|t|^{\frac{3}{2}}+C_6|t|^{\varepsilon},\\
\delta u&\simeq&C_7|t|^{\frac{1}{2}}+C_8|t|^{-1+\varepsilon}.
\eeqn
\end{subequations}
\beqn
\delta U_i &\simeq& C_9, ~~~
{G_i} \simeq  {C_{10}}|t|^{ - 2}. \label{Solution_vector_2} \\
D_{ij}&\simeq& |t|^{-\frac{1}{2}}( C_{11}|t|^{\frac{\sqrt{1+24\varepsilon}}{2}}+ C_{12}|t|^{-\frac{\sqrt{1+24\varepsilon}}{2}}),
\label{Solution_tensor_2}
\eeqn
where $C_i$ are functions of wave vector and $\varepsilon=-\frac{\kappa k^2}{12a_{B}^2}$. Since the scalar $E$, vector $G_i$, and tensor $D_{ij}$ diverge
near the maximum density ($t\rightarrow0$), we conclude that the scalar, vector, and tensor modes will all cause instabilities in the Eddington regime in this case.

\section{Conclusions}

In this paper, we study the full linear perturbations in the radiation era of a homogeneous and isotropic spacetime in the EiBI theory. The former research showed that even though the spacetime is singularity free at and early time, the transverse-traceless tensor perturbation could cause an instability in this era \cite{Escamilla-Rivera2012}. We linearize the field equations and solve the full linear perturbed modes with the approximate background solutions near the maximum density. Interestingly, we find that, for $\kappa>0$, the scalar modes are stable for $k=0$ but linearly unstable for $k\neq0$. This instability was also pointed out in Ref. \cite{Lagos2013}, recently. The vector modes are stable and the tensor mode is unstable in the Eddington regime,  independent of the wave vector $k$. However, for $\kappa<0$, all the scalar, vector, and tensor modes cause instabilities in Eddington regime.

In Einstein theory, the solution of the scale factor in the radiation era is $a(t)=a_0 \sqrt{t}$, and the Friedmann equation is $(\fc{\dot a}{a})^2=\fc{\rho}{3}$. By solving the corresponding linear perturbed equations \cite{Weinberg2008}, the perturbed modes are given as
$E=-A\simeq d_1t^{-\fc{3}{2}}+d_2$, $\delta u \simeq d_1t^{-\fc{1}{2}}-\fc{d_2}{2}t$, ${\delta\rho}\simeq\fc{3}{2}d_1 t^{-\fc{7}{2}}-\fc{3}{4}d_2 t^{-2}$, $G_i\simeq d_3t^{-\fc{1}{2}}$, $\delta {U_i}\simeq d_4 t^{\fc{1}{2}}$, and $D_{ij}\simeq d_5 t^{-\fc{1}{2}} + d_6$, where the $d_i$ are functions of wave vector. It shows that the scalar, vector, and tensor modes are all unstable in the early Universe ($t\rightarrow0$).  Thus, the EiBI cosmology with $\kappa>0$ presents as an interesting theory with stable scalar (when $k=0$) and vector modes. This was also demonstrated in the research \cite{Pani2011,Pani2012,Avelino2012a}, in which the EiBI theory with positive $\kappa$ showed more good properties than negative ones.

\section*{Acknowledgements}
{The authors would like to thank R.G. Cai for helpful discussions.
This work was supported in part by the National Natural Science Foundation of China (Grants No.~11075065 and No.~11375075)
and the Fundamental Research Funds for the Central Universities (Grant No. lzujbky-2013-18). K. Yang was supported by the Scholarship Award for Excellent Doctoral Student granted by Ministry of Education.}

%\section*{References}

%\bibliographystyle{unsrt}
%\bibliographystyle{JHEP}
%\bibliographystyle{apsrev}
%\bibliographystyle{apsrmp4-1}
%\bibliography{E:/1_Physics/1_Work/4_Other/JbRef_DataBase/Articles,E:/1_Physics/1_Work/4_Other/JbRef_DataBase/Books}

\end{document}